# The BlueWalker 3 Satellite Has Faded

Anthony Mallama, Richard E. Cole and Scott Tilley

2022 December 26

Contact: anthony.mallama@gmail.com

Abstract

Observations of BlueWalker 3 (BW3) beginning on December 8 of this year indicate that its apparent brightness had decreased. We postulate that the orbital beta angle and resultant solar power considerations required an adjustment to the satellite attitude around that time. So, the nominally zenith facing side of the flat-panel shaped spacecraft, which supports the solar array, was tilted toward the Sun. Consequently, the nadir side, which is seen by observers on the ground, was mostly dark. Thus, BW3 has generally appeared faint and on some occasions was not seen at all. The amount of fading was up to 4 magnitudes. Numerical modeling indicates that the amount of tilt was in the range 13° to 16°. This situation indicates the improvement in the appearance of BW3 from the ground that can be achieved with small tilts of the spacecraft. Satellite operators and astronomers can jointly address the adverse impact of bright satellites on celestial observations based on this finding.

1. Introduction

BlueWalker 3 (BW3) is the prototype for a new constellation of satellites. Astronomers are concerned because this type of spacecraft unfolds into an extremely large size on-orbit and becomes very bright. The International Astronomical Union issued a [press release](#) regarding the adverse affect of BW3 and its constellation on astronomical research and on the appearance of the night sky. Mroz et al. (2022) and Halferty et al. (2022) present photometric results for other artificial satellites that document their impact on scientific observations.

Mallama et al. (2022) reported that the apparent visual magnitude of BW3 is most often between 2.0 and 3.0. The average brightness when seen near zenith at the beginning and ending of astronomical twilight is magnitude 1.4. Those findings were based on observations made between the time that the spacecraft unfolded and November 24 of this year.

The magnitudes recorded more recently are significantly fainter. This paper documents that fading and suggests an explanation. Section2 describes a hypothesis for the dimming which relates to the spacecraft attitude.  Section3 summarizes recent observations when BW3 was faint or was not seen at all. Section 4 compares the observations to a model of the satellite's brightness that accounts for the attitude. Section 5 presents the conclusions and suggests how the adverse impact of BW3 on astronomical observations can be ameliorated.

2. Spacecraft attitude and orbital beta angle

BW3 unfolded into a 64 square meter flat-panel antenna on orbit. The nominal spacecraft attitude (yaw, pitch and roll angles) has the two sides of the panel facing the zenith and nadir directions. The nadir side contains radio communication elements while the zenith side supports the solar power array. During the course of an orbit around the Earth, the solar array is exposed to sunlight about half the time.

In December of this year the satellite orbital plane became nearly perpendicular to the direction of the Sun. The angle between the orbital plane and the solar direction is called *beta* (Versteeg and Cotton, and Sumanth 2019). When the cosine of beta is small a zenith-nadir facing panel is almost edge-on to the Sun throughout the orbit. Thus, solar power is severely curtailed unless the spacecraft attitude is adjusted. The value of beta for BW3 and its cosine are plotted versus time in Figure 1.

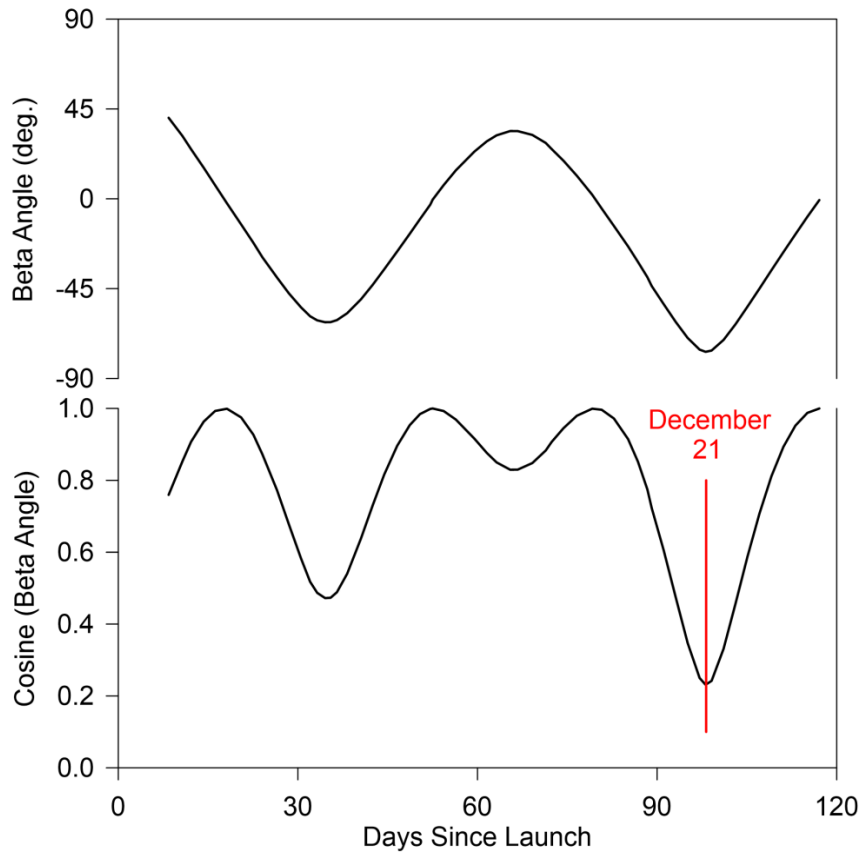

*Figure 1. The beta angle and its cosine are illustrated, and the minimum in the cosine value on December 21 is indicated.*

We hypothesize that the spacecraft attitude has been adjusted by tilting the nominally zenith-facing solar array toward the Sun. Figure 2 illustrates the tilt which is mainly a roll angle adjustment. Tilting the panel in this manner increases insolation on the array and, as a consequence, darkens the nadir side that faces observers on the ground. Thus, sunlight strikes the nadir side from a direction that is more parallel to the plane of the panel than usual. This reduces illumination of the panel and less light is reflected toward the ground. So, the satellite will appear fainter. Additionally, sunlight may strike the zenith side only, in which case the satellite will be invisible or nearly invisible from the ground. In the 'nearly invisible' case, some light may reflect from the edge of the panel and a small amount may leak through the interstices between the articulated panel elements.

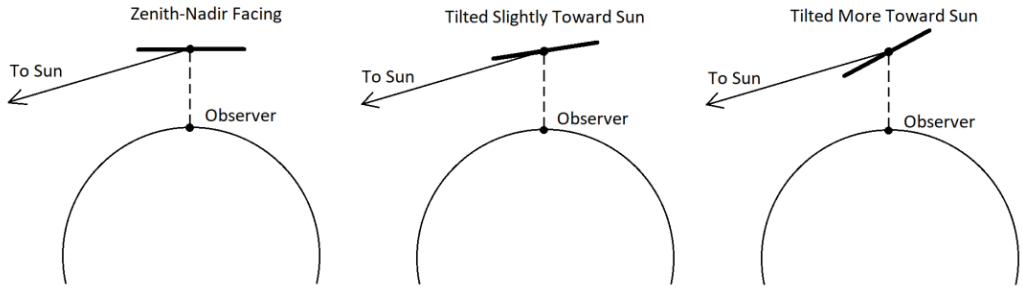

*Figure 2. (Left) This schematic shows the nominal spacecraft attitude where the flat-panel is zenith-and-nadir facing. There is no insolation on the solar array in that case. (Middle) This is an adjusted attitude where the zenith facing side of the panel is tilted slightly toward the Sun and the nadir side receives less insolation. (Right) The panel is tilted further. So, no sunlight reaches the nadir side and the satellite is invisible. (All) This Sun-satellite-observer geometry would apply to a satellite seen near zenith at about the time when astronomical twilight begins or ends. The beta angle is large but not exactly 90 degrees in these schematics.*

3. Observations

The magnitudes used to study BW3 were obtained by the observers listed in Table 1. Most of the measurements were made with the unaided eye or through binoculars. Visual magnitudes are determined by comparing the brightness of the satellite to that of nearby reference stars. This proximity accounts for variations in sky transparency and brightness. The method is described in more detail by Mallama (2022). Some of the observations were derived from video recordings by Langbroek. He transformed the red-sensitive magnitudes to the V-band using an empirical formula based on the analysis of reference star measurements.

Table 1. Observer coordinates

| Observer | Latitude | Longitude | Ht(m) |
|---|---|---|---|
| J. Barentine | 32.234 | -110.768 | 833 |
| R. Cole | 50.552 | -4.735 | 100 |
| K. Fetter | 44.606 | -75.691 | |
| S. Harrington | 36.062 | -91.688 | 185 |
| M. Langbroek | 52.154 | 4.491 | 0 |
| M. Langbroek | 52.139 | 4.499 | -2 |
| R. Lee | 38.93 | 104.81 | 2082 |
| P. Maley | 33.811 | -111.952 | 654 |
| P. Maley | 32.857 | -113.220 | |
| P. Maley | 34.6 | 33.0 | 0 |
| A. Mallama | 38.982 | -76.763 | 43 |
| A. Mallama | 38.72 | -75.08 | 0 |
| A. Mallama | 39.122 | -77.891 | |
| R. McNaught | -32.27 | 149.16 | 610 |
| J. Respler | 40.330 | -74.445 | |
| R. Swaney | 41.403 | -81.512 | |

```
              S. Tilley           49.434    -123.668     40
              S. Tilley           49.418    -123.642      1
              E. Visser           53.109       6.108     46
              A. Worley           41.474     -81.519    351
              J. Worley           41.474     -81.519    351
              B. Young            36.139     -95.983    201
              B. Young            35.831     -96.141    330
```

Figure 3 shows the light curve for BW3 after the satellite unfolded its large antenna panel. The magnitudes are adjusted to a standard distance of 1000 km. The average brightness before the fading that began around December 8 was magnitude 3.05. These data are referred to as *regular brightness* in the graph. The average observed brightness during the *fainter period* is magnitude 5.72 and that does not include the times when the satellite was too faint to be seen.

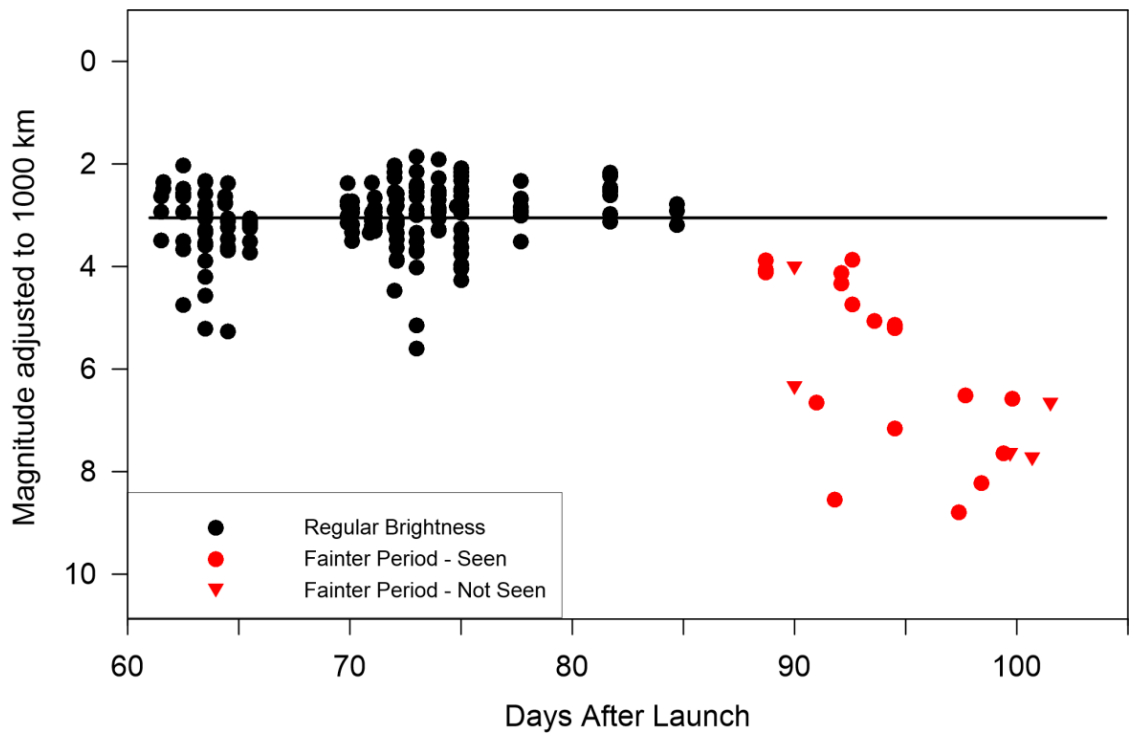

*Figure 3. The light curve of BW3 shows that observations beginning on December 8 are all fainter than the average magnitude before that date. The 'not seen' symbols indicate the magnitude of the faintest visible comparison star; so the satellite was fainter than that value.*

4. Comparison between observations and model

Figure 4 shows the BW3 spacecraft before launch. The antenna panel which can be seen from the ground is visible. The solar panels are on the other face, but that face cannot be seen from the ground. The antenna panel appears in the image as a diffusely reflecting surface, there are no significant areas of specularly reflecting materials as there are on the Starlink spacecraft, both in the original Starlink design and later updates.

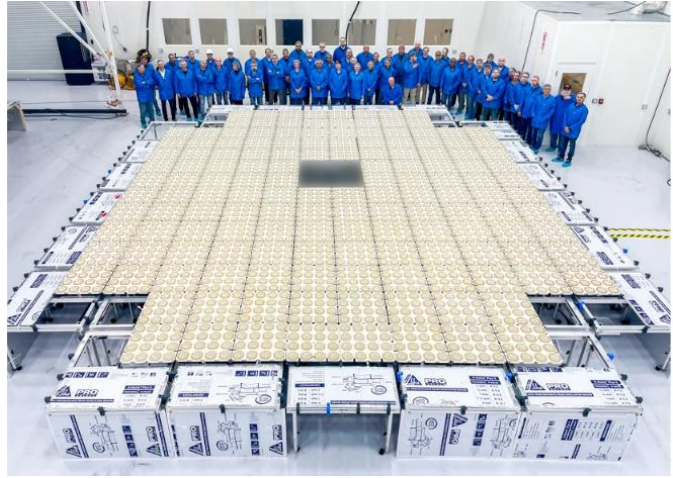

*Figure 4. The nadir face of the BlueWalker spacecraft during a pre-flight test at Cape Canaveral. The general appearance is of a diffusely reflecting flat panel (photo courtesy AST).*

A numerical model for the brightness of BW3 was developed, building on experience from observing and modeling Starlink spacecraft since 2020 (Cole 2020, 2021).

In the case of BW3, the numerical model uses a single Earth-facing flat surface that reflects light diffusely. This diffuse reflection is well described by Lambert's cosine law. More complex models of reflection from the panel are available but not required to analyze the 'fading' discussed here.

The model takes account of the aspect angle of the panel with respect to the observer, its range and the angle of the Sun illumination on the panel (which is different from the angle of the Sun at the observer).

The model was developed using observations made after full deployment of BW3. In that case the panel was maintained in the model as nadir-facing and the only variable was an absolute brightness of the panel. With suitable selection of that absolute brightness, a reasonable match was found between the predictions of the model and the visual observations (Figure 5).

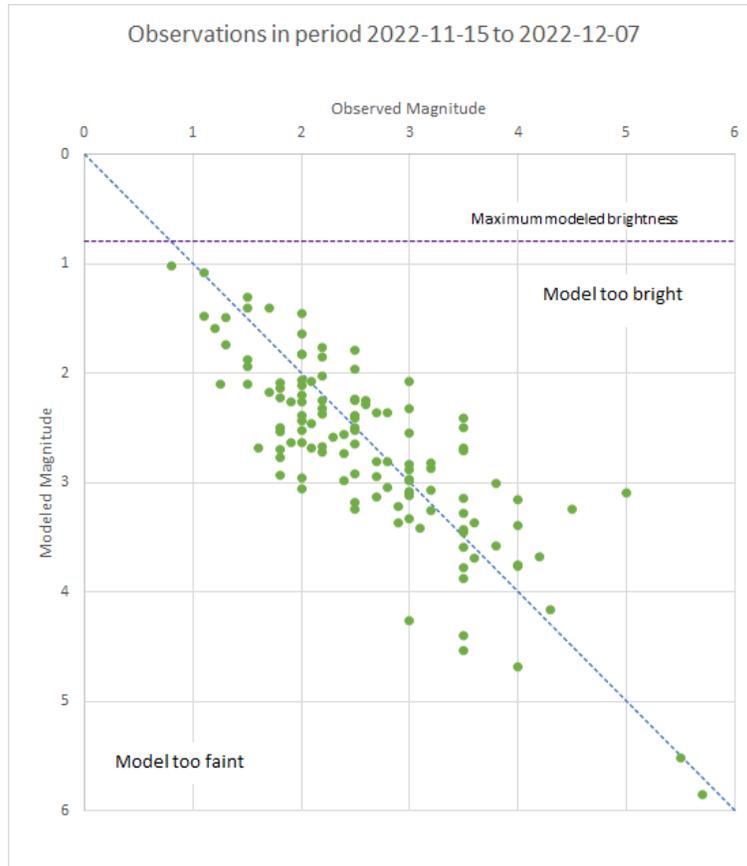

*Figure 5. Comparison of the brightness predictions of the numerical model and observations made in the period after the AST announcement of full deployment and before December 7. The closer the points are to the diagonal line, the closer the model is to the observed brightness.*

The maximum brightness of BW3 is always in the zenith and is a strong function of the elevation of the Sun, brighter when the Sun is further below the horizon at the observer. The brightest observations were made when BW3 was in the zenith and entering or emerging from eclipse.

However, from December 8 of this year observations of BW3 were not well described by the same model. The same comparison of model and observations (as in Figure 5) is displayed in Figure 6-i. Cases where BW3 was below the limiting magnitude of the observation are shown with an arrow.

Clearly, the model no longer predicts the BW3 brightness to any level of accuracy. Some of the observations are 3 or 4 magnitudes fainter than their prediction, or BW3 was not seen at all. Since the surface material of the BW3 panel has not changed, the most likely explanation was a change in the spacecraft attitude for the reasons discussed above, with a movement of the zenith axis towards the Sun. A tilt of this sort reduces the angle of the Sun on the Earth-facing panel and reduces the amount of sunlight that is scattered, thus making BW3 fainter (Figure 2).

A tilt was added as a parameter of the model and a range of tilt angles investigated. A best fit of the predictions to the data was achieved for a tilt of 13° but tilts up to 16° are a reasonable fit to the data (Figure 6-ii). Tilts of less than 13° do not fit the data. With this deduction, a small number of the observations had been made when the Earth-facing panel was not illuminated by the Sun at all, that is the Sun was shining on the solar-panel side. In these cases, the BW3 brightness was at the limit of observations using binoculars, magnitudes fainter than 7 or 8. It is hypothesized that this limited solar flux is being reflected from the edges of the panel or the joints between the panel elements. A simple model was added that fitted this small number of observations.

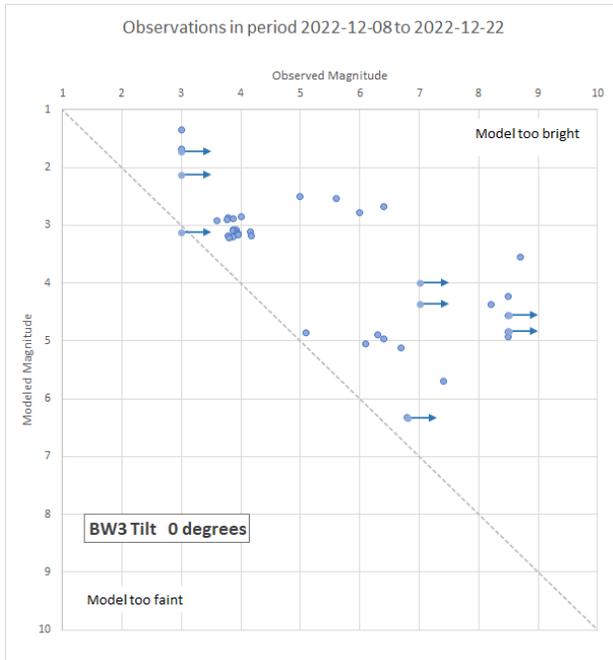
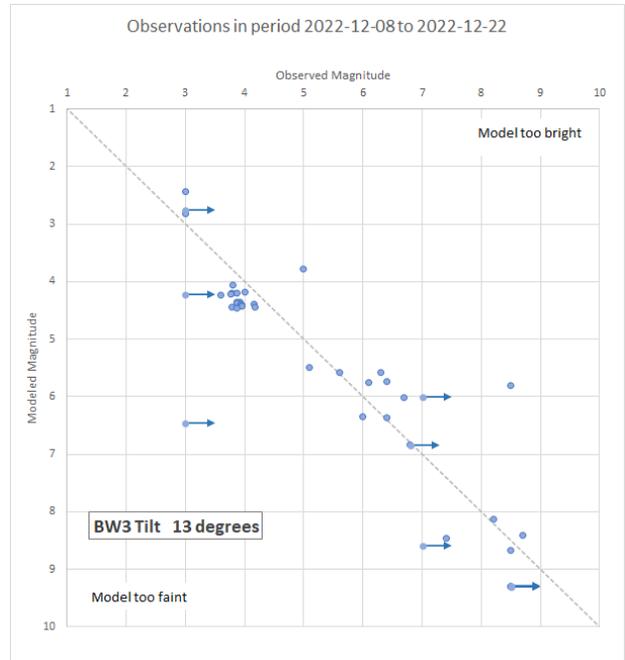

*Figure 6 i) Comparison of the brightness predictions of the numerical model and the observations taken in the period December 8 to 22, using the same model parameters as in Figure 5.*

*Figure 6 ii) The same data using a model that tilts the BW3 solar panel 13° towards the Sun. A simple model has been added to deal with cases when the Sun is only illuminating the zenith side of the panel.*

Using this model, brightness maps can be created for the whole sky. Individual maps are required for each elevation of the Sun at the observer as this affects the appearance of the satellite across the whole sky. Figure 7 displays polar projection maps for Sun elevations of -18° (end of astronomical twilight) and -12° (end of nautical twilight), in each case with no tilt and with a tilt of 13°. The azimuth of the Sun has been standardized to 90°. The following can be noted:

1. In all the maps BW3 is in eclipse over part of the sky opposite the Sun – the anti-Sun direction (white areas)
2. In the untilted skymaps (left of Figure 7) BW3 is brighter when the Sun is further below the horizon due to the greater angle of the Sun on the Earth-facing antenna panel. Otherwise, the appearance of BW3 is similar for the two Sun elevation cases.

3. In the tilted skymaps (right of Figure 7) the appearance of BW3 is more complex and also different for the two Sun elevations. In the pro-Sun direction the observer sees the un-illuminated antenna panel and thus BW3 is faint, below 7th magnitude. In the anti-Sun direction BW3 is much fainter than the untilted cases, with the effect more pronounced when the Sun elevation is -12°. This is because a tilt of 13° very significantly reduces the Sun illumination on the antenna panel across the whole sky.

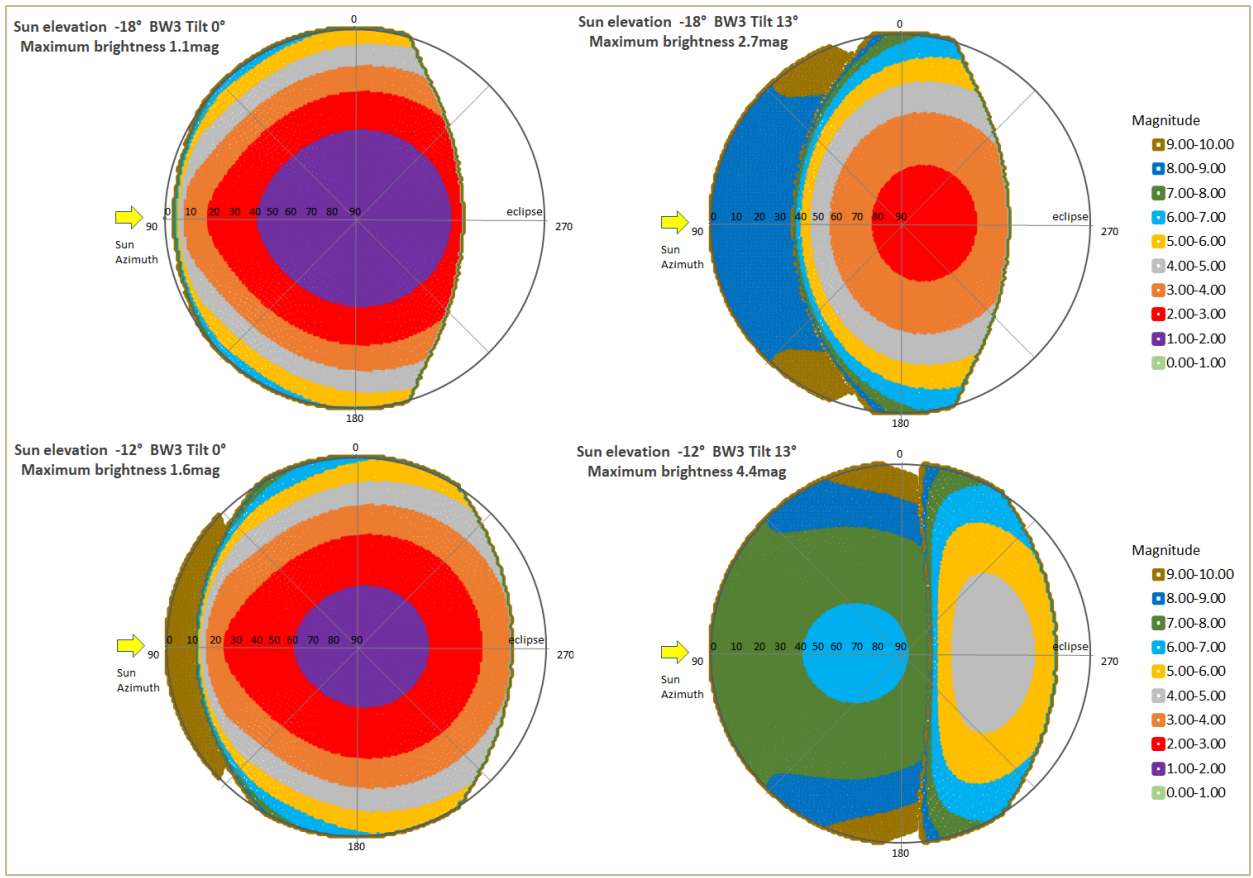

*Figure 7: Skymaps of BW3 brightness for sun elevations of -18° and -12°, no tilt (left) and with a sunward tilt of 13° (right). A polar projection is used, north at the top and west at the right. The Sun azimuth of 90° is shown here but the relative appearance is the same for any Sun azimuth.*

## 5. Conclusions

We offer a hypothesis to explain the observed fading of the BlueWalker 3 satellite that began on December 8 of this year. The orbital beta angle at that time resulted in low insolation on the zenith facing side of the spacecraft which supports the solar array. Power considerations required an adjustment to the satellite attitude such that the zenith side was tilted toward the Sun to increase insolation. Consequently, the nadir side of the panel seen by observers on the ground was dark or only weakly illuminated by the Sun.

While this situation appears to have arisen for operational reasons, it demonstrates that even a small change of spacecraft attitude has a major effect on the brightness of BlueWalker 3 as viewed from the ground, though this effect is less pronounced when the Sun is further below the horizon. Satellite operators and astronomers can begin a constructive dialog based on this finding.

References

Cole, R.E. 2020. A sky brightness model for the Starlink 'Visorsat' spacecraft – I, Research Notes of the American Astronomical Society, 4, 10, https://iopscience.iop.org/article/10.3847/2515-5172/abc0e9

Cole, R.E. 2021. A sky brightness model for the Starlink 'Visorsat' spacecraft. https://arxiv.org/abs/2107.06026

Halferty, G., Reddy, V., Campbell, T., Battle, A. and Furaro, R. 2022. Photometric characterization and trajectory accuracy of Starlink satellites: implications for ground-based astronomical surveys. https://arxiv.org/abs/2208.03226.

Mallama, A., Cole, R.E., Harrington, S. and Maley, P.D. 2022. Visual magnitude of the BlueWalker 3 satellite. https://arxiv.org/abs/2211.09811.

Mallama, A., 2022. The method of visual satellite photometry. https://arxiv.org/abs/2208.07834.

Mroz, P., Otarola, A., Prince, T.A., Dekany, R., Duev, D.A., Graham, M.J., Groom, S.L., Masci, F.J. and Medford, M.S. 2022. Impact of the SpaceX Starlink satellites on the Zwicky Transient Facility survey observations. https://arxiv.org/abs/2201.05343.

Sumanth, R.M. 2019. Computation of eclipse time for low-Earth orbiting small satellites. https://commons.erau.edu/cgi/viewcontent.cgi?article=1412&context=ijaaa

Versteeg, C. and Cotten, D.L. Preliminary thermal analysis of small satellites. https://s3vi.ndc.nasa.gov/ssri-kb/static/resources/Preliminary_Thermal_Analysis_of_Small_Satellites.pdf